 \documentclass[final,5p,times,twocolumn,authoryear]{elsarticle}

\usepackage{amssymb}
\usepackage{lipsum}
\usepackage{xcolor}
\journal{Astronomy $\&$ Computing}

\begin{document}

\begin{frontmatter}



\title{AI-based separation of turbulence from coherent background flows in decaying hydrodynamic turbulence}


\author[first]{Ji-Hoon Ha}
\affiliation[first]{organization={Korea Astronomy and Space Science Institute},%Department and Organization
            addressline={776, Daedeok-daero, Yuseong-gu}, 
            city={Daejeon},
            postcode={34055}, 
            country={Republic of Korea}}

\author[second]{Elena S. Volnova}
\affiliation[second]{organization={Institute for Basic Science},%Department and Organization
            addressline={55, Expo-ro, Yuseong-gu}, 
            city={Daejeon},
            postcode={34126}, 
            country={Republic of Korea}}

\begin{abstract}
Separating turbulent fluctuations from coherent large-scale background flows is a longstanding challenge in the analysis of numerical simulations and astronomical observations.
Traditional approaches commonly rely on decomposition-based techniques such as Fourier or wavelet filtering, which assume that a meaningful separation can be achieved through scale selection.
In realistic flows, however, coherent motions and turbulence often overlap across a broad range of scales and interact nonlinearly, making a unique separation inherently ambiguous.
In this work, we investigate the robustness of an AI-based turbulence–background separation approach using two-dimensional incompressible Navier–Stokes simulations of decaying hydrodynamic turbulence.
The simulations are initialized with a coherent background flow and divergence-free turbulent perturbations with a Kolmogorov-like spectrum and evolve without external forcing, providing a controlled physical testbed.
A neural network trained exclusively on static synthetic images is applied to simulation snapshots at different evolutionary stages.
The model recovers turbulent fluctuations during early and intermediate stages when partial scale separation is present.
At later stages, where nonlinear interactions increasingly mix coherent and turbulent structures, the separation becomes less distinct; nevertheless, the recovered fields remain visually and spectrally consistent with the expected turbulent behavior.
Quantitative comparisons with a Fourier filtering baseline show that the AI-based approach achieves comparable reconstruction accuracy while not requiring an explicit spectral cutoff scale.
These results suggest that AI models trained on static data can provide a flexible diagnostic tool for turbulence–background separation in time-evolving flows, with potential applications to astrophysical datasets.
\end{abstract}



\begin{keyword}
%% keywords here, in the form: keyword \sep keyword, up to a maximum of 6 keywords
Background separation \sep Deep learning \sep Hydrodynamic simulations \sep Turbulence



\end{keyword}

\end{frontmatter}

%\tableofcontents



\section{Introduction}
\label{sec:s1}

Turbulent flows in natural and astrophysical environments are seldom encountered in isolation.
Turbulent fluctuations are generally embedded within coherent large-scale background flows produced by global shear, stratification, rotation, or other organized motions.
For instance, velocity fields in molecular clouds display pronounced large-scale gradients alongside small-scale turbulent fluctuations \citep[e.g.,][]{Imara2011, Li2021}.
Similarly, magnetic fields in molecular clouds consist of smoothly varying large-scale components superimposed with turbulence on smaller scales \citep[e.g.,][] {Girart2006,Hildebrand2009,Houde2009,Pattle2017,Pattle2022}.
Separating turbulent fluctuations from these background components has therefore remained a long-standing challenge in the analysis and interpretation of both observational data and numerical simulations.

\begin{figure*}[t]
    \centering
    \includegraphics[width=1\linewidth]{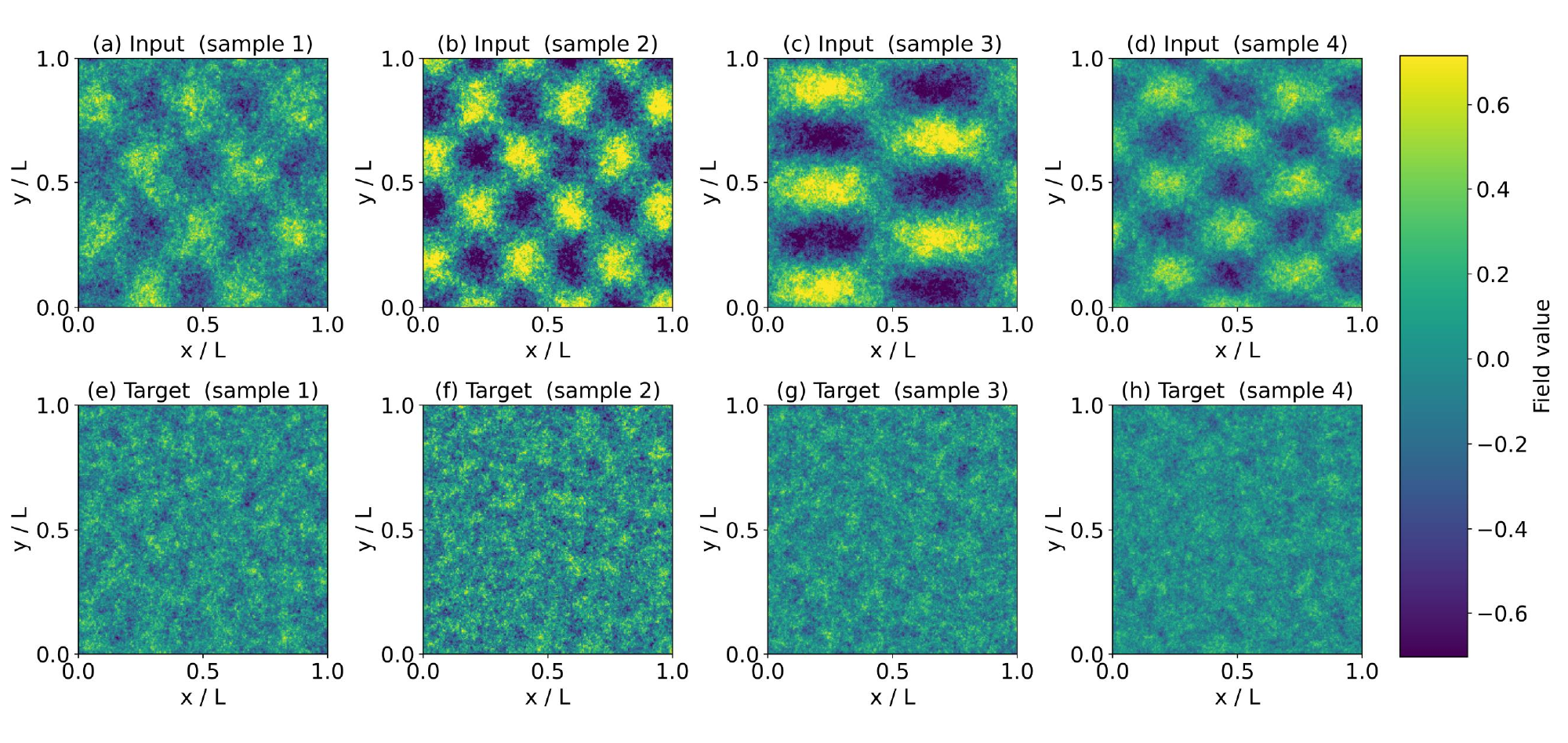}
    \caption{Representative examples of the synthetic training data used for the AI-based turbulence--background separation. The top row shows input images constructed as linear superpositions of a coherent large-scale background field and a turbulent fluctuation field ($O(x,y) = B(x,y) + I(x,y)$). The bottom row shows the corresponding turbulence-only targets ($I(x,y)$). The background component is dominated by low-wavenumber coherent structures, while the turbulent component exhibits small-scale, nearly isotropic fluctuations with a Kolmogorov-like power spectrum.}
    \label{fig:f1}
\end{figure*}

Traditional approaches to separating turbulent fluctuations from large-scale background flows commonly rely on decomposition-based techniques, including scale-based filtering methods such as Fourier or wavelet transforms \citep[e.g.,][]{Grebenev1995}, as well as adaptive methods like the Hilbert--Huang transformation \citep{Huang1998,Huang1999}.  
In these approaches, a cutoff scale or a selected set of modes is typically used to distinguish coherent background variations from turbulent fluctuations.  
While such methods are effective and widely employed, they inevitably involve subjective choices and implicitly assume that a physically meaningful separation can be achieved through scale or mode selection alone.  
More sophisticated multiscale statistical frameworks incorporating higher-order moments and information-theoretic measures have been developed to characterize turbulence beyond second-order statistics. However, these approaches likewise rely on multiscale decompositions and require an explicit or implicit removal of large-scale trends, rendering their interpretation sensitive to how background components are defined and subtracted \citep{GraneroBelinchon2024}.  
In realistic flows, coherent motions and turbulent fluctuations often coexist and overlap across a broad range of scales and interact nonlinearly, making any clear and unique separation based solely on scale inherently ambiguous \citep[e.g.,][]{Cho2019}.  
This ambiguity is further compounded in astrophysical applications, where observables are frequently projected along the line of sight and integrate contributions from multiple scales with comparable amplitudes, complicating attempts to disentangle coherent and turbulent components even when their physical scales are well motivated \citep{SetaFederrath2024}.  
Moreover, the definition of a local, scale-dependent background field itself can influence measured statistics and intermittency, highlighting that the distinction between background and fluctuations is not uniquely defined in multiscale, anisotropic turbulent systems \citep{Sioulas2025}.

Recent advances in deep learning have opened new possibilities for turbulence–background separation by learning non-local and multi-scale features directly from data \citep[e.g.,][]{Brunton2020}. 
Several studies have demonstrated that neural networks can recover turbulent components with high fidelity under controlled conditions when trained on synthetic datasets constructed by linear superposition of independently generated background and turbulence fields
\citep[e.g.,][]{Ling2016, Nakamura2021}. 
Despite these promising results, an important open question remains: whether models trained on such static and idealized data can generalize to physically realistic systems in which background flows and turbulence evolve self-consistently and interact through nonlinear dynamics \citep[e.g.,][]{Brenner2019, Kochkov2021}.

In this work, we address this question by testing an AI-based turbulence--background separation model in a time-evolving hydrodynamic setting.
Specifically, we apply a neural network trained exclusively on synthetic static fields constructed from linear superpositions of background flows and turbulent fluctuations to snapshots from two-dimensional incompressible Navier–Stokes simulations of decaying hydrodynamic turbulence (see Sec. 3.1 for details of the simulation setup).
The simulations are initialized with a coherent large-scale background flow and divergence-free turbulent perturbations with a Kolmogorov-like power spectrum, and subsequently evolve without external forcing.
As the turbulent energy decays and nonlinear interactions progressively distort the background flow, the distinction between coherent and turbulent structures becomes increasingly ambiguous, providing a conservative and physically meaningful stress test for AI-based separation.
Rather than proposing a new separation algorithm, the present study uses AI as a diagnostic tool to probe the physical conditions under which turbulence–background separation remains meaningful in time-evolving flows.

The goal of this study is not to construct a fully realistic model of astrophysical turbulence, but rather to assess the robustness and potential of AI-based turbulence--background separation under controlled yet physically nontrivial conditions.
By systematically applying the trained model to simulation snapshots at different evolutionary stages, we aim to identify the regime in which such separation remains meaningful, as well as the conditions under which it breaks down.
The paper is organized as follows.
In Section \ref{sec:s2}, we describe the AI model and the construction of the synthetic training data.
Section \ref{sec:s3} presents the hydrodynamic simulations and the application of the AI model to decaying turbulent flows.
Summary and discussion are provided in Section \ref{sec:s4}.

\section{AI Model for Turbulence--Background Separation}
\label{sec:s2}

\subsection{Problem Definition and Training Data}
The training dataset consists of synthetic two-dimensional fields constructed by linearly superposing a coherent background flow and a turbulent fluctuation component.
We formulate turbulence--background separation as an image-to-image regression task.
Given a 2D input field $O(x,y)$ that contains a coherent large-scale background component
$B(x,y)$ and a turbulent fluctuation field $I(x,y)$,
\begin{equation}
O(x,y) = B(x,y) + I(x,y),
\end{equation}
the goal is to infer the turbulent component $\hat{I}(x,y)$ from the input.
For training, we construct synthetic examples where the background and turbulence are
generated independently and then linearly superposed.
The background component is modeled as a low-wavenumber sinusoidal pattern,
\begin{equation}
B(x,y) = A\,\sin\!\left(2\pi k_x x + \phi_1\right)\cos\!\left(2\pi k_y y + \phi_2\right),
\end{equation}
with random phases $(\phi_1,\phi_2)$ and randomized spatial frequencies
$(k_x,k_y)$ sampled uniformly in the range corresponding to $1$--$3$ cycles across the
$224\times224$ domain. The amplitude $A$ is randomly sampled in the range
$[0.3,\,0.8]$ per sample. This construction follows the implementation in our
dataset generator.
Turbulent fields are generated as isotropic random-phase realizations with a
Kolmogorov-like power spectrum,
\begin{equation}
P_I(k)\propto k^{-5/3},
\end{equation}
implemented by assigning Fourier amplitudes $|\hat{I}(\mathbf{k})|\propto k^{-5/6}$
above a prescribed injection wavenumber and then transforming back to real space.
The injection scale is fixed to $\ell_{\rm inj} = L/8$ (i.e., an injection scale ratio of $1/8$),
and the root-mean-square (RMS) amplitude of the turbulent field is randomized in the range $[0.05,\,0.25]$ for each sample. This randomization introduces variations in the overall turbulent energy across the training dataset, exposing the model to a range of fluctuation amplitudes and reducing sensitivity to a specific energy normalization.

The turbulence--background separation problem is formulated as an image-to-image regression task. 
The network input consists of a two-dimensional scalar field $O(x,y)$ that contains both a coherent large-scale background component and small-scale turbulent fluctuations. 
The target output is the turbulent component $I(x,y)$ alone.
Since the data are intrinsically single-channel scalar fields, both the input and the target are represented as tensors of shape $(1,224,224)$.

Fig. \ref{fig:f1} shows representative examples of the synthetic training data.
The top row displays input images constructed by linearly superposing a smooth background field and a turbulent field, while the bottom row shows the corresponding turbulence-only targets. The background component is dominated by low-wavenumber modes with coherent spatial structure, whereas the turbulent component exhibits nearly isotropic small-scale fluctuations with a Kolmogorov-like spectrum.

\subsection{Network Architecture}

\begin{figure}[t]
    \centering
    \includegraphics[width=1\linewidth]{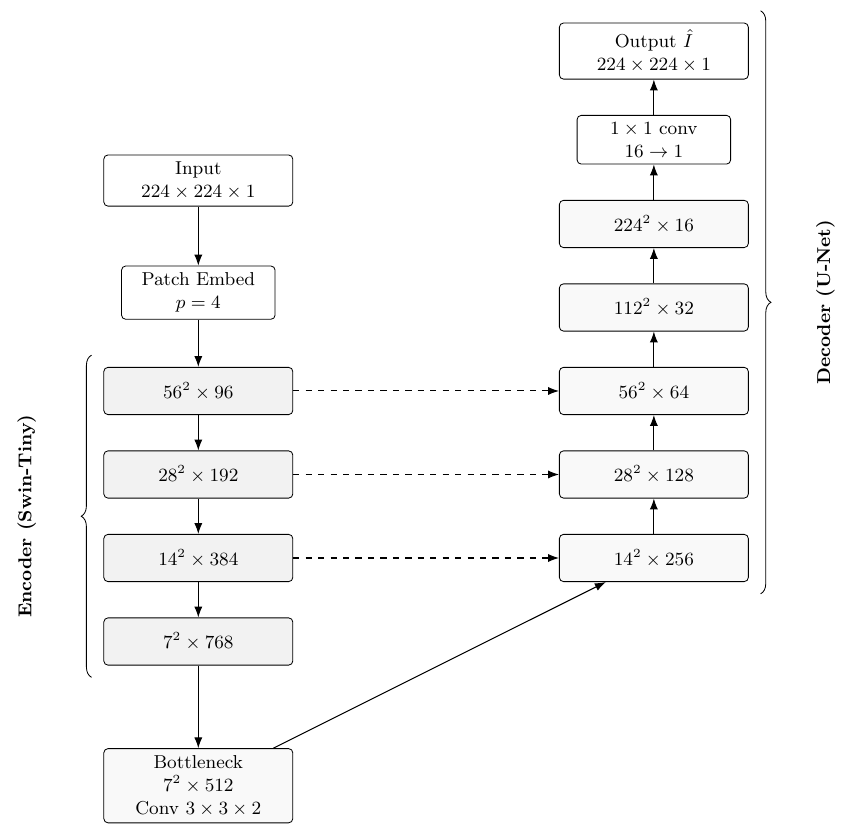}
    \caption{Architecture of the hybrid Swin Transformer–U-Net model used for turbulence–background separation. The Swin-Tiny encoder extracts hierarchical multi-scale features at resolutions $56^2$, $28^2$, $14^2$, and $7^2$, which are passed to a U-Net–style decoder via same-resolution skip connections. The decoder progressively upsamples the latent representation and reconstructs the target field at the original resolution. Both the input and the output are single-channel scalar fields of size $224\times224$.}
    \label{fig:f2}
\end{figure}

Our separation model is a hybrid Transformer--CNN U-Net designed for multi-scale
reconstruction.
The encoder is a Swin Transformer (Swin-Tiny, patch size $4$, window size $7$) implemented
via \texttt{timm}, and multi-scale feature maps are extracted at four resolutions
corresponding to patch-embedding output and successive Swin stages \citep[][]{Liu2021}. A schematic overview of the model architecture is shown in Fig. \ref{fig:f2}. 
The encoder is a Swin Transformer that operates on non-overlapping image patches and captures long-range spatial correlations through shifted self-attention windows.
Multi-scale feature maps extracted at successive encoder stages are passed to the decoder through skip connections.
The decoder progressively upsamples these features and reconstructs the turbulent component at the original spatial resolution \citep[][]{Ronneberger2015}.
This design enables the model to distinguish between large-scale coherent patterns and small-scale turbulent structures.

Specifically, for an input image of size $224\times224$, the encoder produces feature maps at resolutions $56^2$, $28^2$, $14^2$, and $7^2$ with nominal channel dimensions
$\{96,192,384,768\}$ for Swin-Tiny. 
These are projected via $1\times1$ convolutions to a U-Net-like set of skip features with channels $\{64,128,256,512\}$.
The decoder follows a U-Net design with bilinear upsampling, skip concatenation, and two-layer $3\times3$ convolutional blocks:
\begin{itemize}
\item Bottleneck: a convolutional block operating at $7\times7$ and 512 channels.
\item Upsampling path: three upsampling blocks with skip connections
($7\rightarrow14\rightarrow28\rightarrow56$).
\item Final refinement: two additional upsampling steps without skip connections
($56\rightarrow112\rightarrow224$) using convolutional blocks (64$\rightarrow$32 and 32$\rightarrow$16).
\item Output head: a $1\times1$ convolution mapping 16 channels to the 1-channel prediction.
\end{itemize}
This architecture outputs $\hat{I}$ at the original resolution, enabling direct comparison to the turbulence target in pixel space. The model contains approximately 35.9 million trainable parameters. The adopted architecture provides a stable baseline configuration, and its sensitivity to moderate variations in the decoder design is examined in \ref{sec:sa1}.

\begin{figure}[t]
    \centering
    \includegraphics[width=1\linewidth]{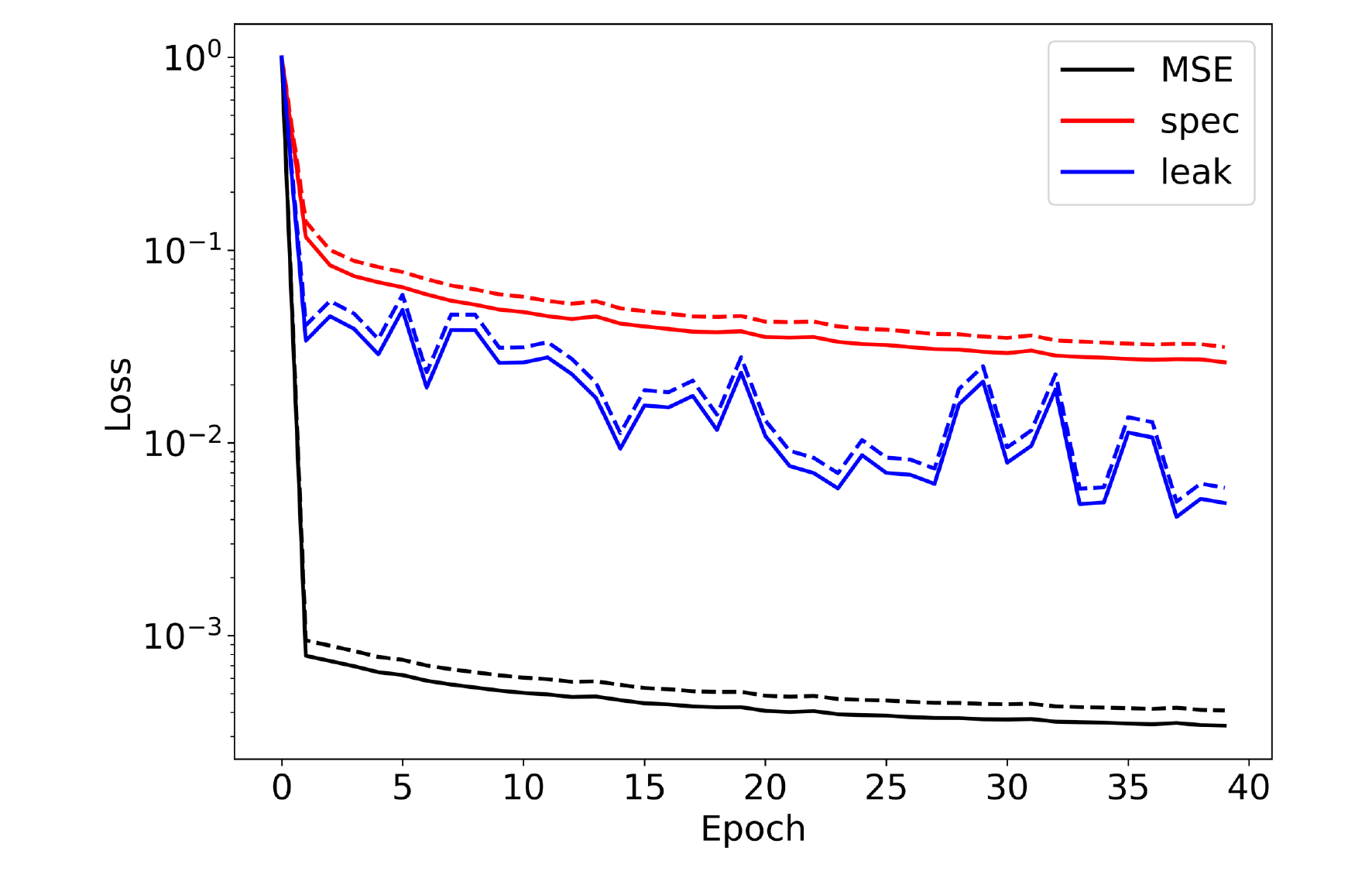}
    \caption{Evolution of the loss components during training. The pixel-wise MSE, spectral consistency loss (spec), and leakage penalty (leak) are shown as functions of epoch. Solid lines represent training losses and dashed lines represent validation losses.}
    \label{fig:f3}
\end{figure}

\begin{table}
\begin{tabular}{l c c c} 
 \hline
Model & $\mathcal{L}_{\rm MSE}$ & $\mathcal{L}_{\rm spec}$ & $\mathcal{L}_{\rm leak}$ \\ 
 \hline
 baseline & 	$3.4 \times 10^{-4}$ & $2.6 \times 10^{-2}$ & $4.9 \times 10^{-3}$ \\ 
 $\lambda_{\rm spec} = 0$& $5.1 \times 10^{-4}$ & $4.3 \times 10^{-2}$ & $5.8 \times 10^{-3}$ 	 \\ 
 $\lambda_{\rm leak} = 0$& $4.9 \times 10^{-4}$ & $4.5 \times 10^{-2}$ & $5.2 \times 10^{-3}$ 	 \\
 \hline
\end{tabular}
\caption{Ablation test for the composite loss function on the synthetic test dataset.
The baseline model uses the full loss, including the pixel-wise MSE term, the spectral consistency term, and the leakage penalty.
Setting either $\lambda_{\rm spec}=0$ or $\lambda_{\rm leak}=0$ degrades all evaluation metrics, indicating that both auxiliary terms contribute to stable and accurate turbulence--background separation.}
\label{Table1}
\end{table}

\subsection{Loss Function}

Training minimizes a composite objective designed to promote both pixel-level fidelity
and physically consistent spectral content:
\begin{equation}
\mathcal{L} = \lambda_{\rm MSE}\,\mathcal{L}_{\rm MSE} + \lambda_{\rm spec}\,\mathcal{L}_{\rm spec}
+ \lambda_{\rm leak}\,\mathcal{L}_{\rm leak}.
\end{equation}

The primary term is the mean squared error,
\begin{equation}
\mathcal{L}_{\rm MSE} = \left\langle \left(\hat{I}-I\right)^2 \right\rangle,
\end{equation}
computed over all pixels, channels, and batch elements.

To enforce spectral agreement, we add a radial power spectral density (PSD) matching term.
After subtracting the spatial mean to remove the zero-wavenumber component, we compute 2D FFTs using orthonormal
normalization and evaluate the radially averaged PSD $P(k)$ by binning $|\hat{F}|^2$ on the
unshifted FFT grid.
We then minimize the mean squared difference of $\log_{10}P(k)$ within a selected wavenumber band:
\begin{equation}
\mathcal{L}_{\rm spec}
=
\left\langle
\left[\log_{10} P_{\hat{I}}(k) - \log_{10} P_{I}(k)\right]^2
\right\rangle_{k_{\min}\le k \le k_{\max}},
\end{equation}
where we set $k_{\min}=k_{\rm inj}$ and $k_{\max}=0.35$ (in units of cycles per pixel). The upper wavenumber limit is chosen to exclude modes near the Nyquist scale ($k_{\rm Nyquist}=0.5$ cycles per pixel), where numerical dissipation and discretization effects become increasingly significant.

Finally, to discourage contamination of the prediction by the coherent background modes,
we introduce a leakage penalty based on the projection coefficient of $\hat{I}$ onto the
known background field $B$ (available in the synthetic training set):
\begin{equation}
a = \frac{\langle \hat{I}\,B \rangle}{\langle B^2\rangle}, \qquad
\mathcal{L}_{\rm leak} = \langle a^2\rangle,
\end{equation}
where $\langle\cdot\rangle$ denotes spatial averaging (and additionally averaging over batch and channels). 
Although the leakage term relies on knowledge of the background field and is therefore applicable only during training on synthetic data, it serves to guide the network toward a physically meaningful separation that generalizes to unseen simulation data.

In our baseline configuration, we set $\lambda_{\rm MSE}=1$, $\lambda_{\rm spec}=5\times10^{-3}$ and
$\lambda_{\rm leak}=5\times10^{-2}$.
The coefficients of the spectral and leakage terms were chosen empirically so that the pixel-wise MSE remains the dominant optimization objective, while the additional terms act as regularizers that encourage spectral consistency and suppress contamination from background modes. 
These values were selected so that the different loss components contribute at comparable magnitudes during training despite their intrinsically different numerical scales.
The evolution of the individual loss terms during training is shown in Fig. \ref{fig:f3}.
The mean squared error rapidly decreases within the first few epochs, indicating fast convergence in pixel space.
The spectral loss decreases more gradually, reflecting the increased difficulty of matching the power spectrum across spatial scales.
The leakage loss remains subdominant but non-negligible, demonstrating that the model actively suppresses contamination from large-scale background modes.

To assess the role of the auxiliary loss terms, we performed a simple ablation study in which either the spectral consistency term or the leakage penalty was removed from the training objective.
The results are summarized in Table~\ref{Table1}.
The baseline model, which uses the full composite loss, yields the best overall performance.
Removing either the spectral term or the leakage term increases not only the corresponding loss component but also the pixel-wise reconstruction error.
This indicates that both auxiliary terms act as effective regularizers and help stabilize the separation by enforcing spectral fidelity and suppressing contamination from background modes.

\subsection{Model Training and Results}

We train the model using AdamW optimizer \citep[][]{Loshchilov2019} with a learning rate of $10^{-3}$ for 40 epochs, with batch size 32 on a dataset of 1024 synthetic images.
Although no explicit early stopping criterion was used, the validation loss was monitored during training and was found to approach a plateau after approximately 30 epochs (see Fig. \ref{fig:f3}), indicating that the model had effectively converged.
In addition, we reserve 128 images for validation during training to monitor convergence and prevent overfitting, and a separate set of 128 images is used for testing and quantitative evaluation.
The trained weights are stored for later inference on hydrodynamic simulation snapshots, which represent a more challenging regime because the background and turbulence evolve self-consistently and interact nonlinearly over time.
Although the nominal training set contains 1024 samples, each sample is procedurally generated with randomized phases, amplitudes, spatial frequencies, and independent turbulence realizations. This procedure effectively produces a large diversity of input configurations, reducing the likelihood of simple memorization during training.

\begin{figure*}[t]
    \centering
    \includegraphics[width=0.7\linewidth]{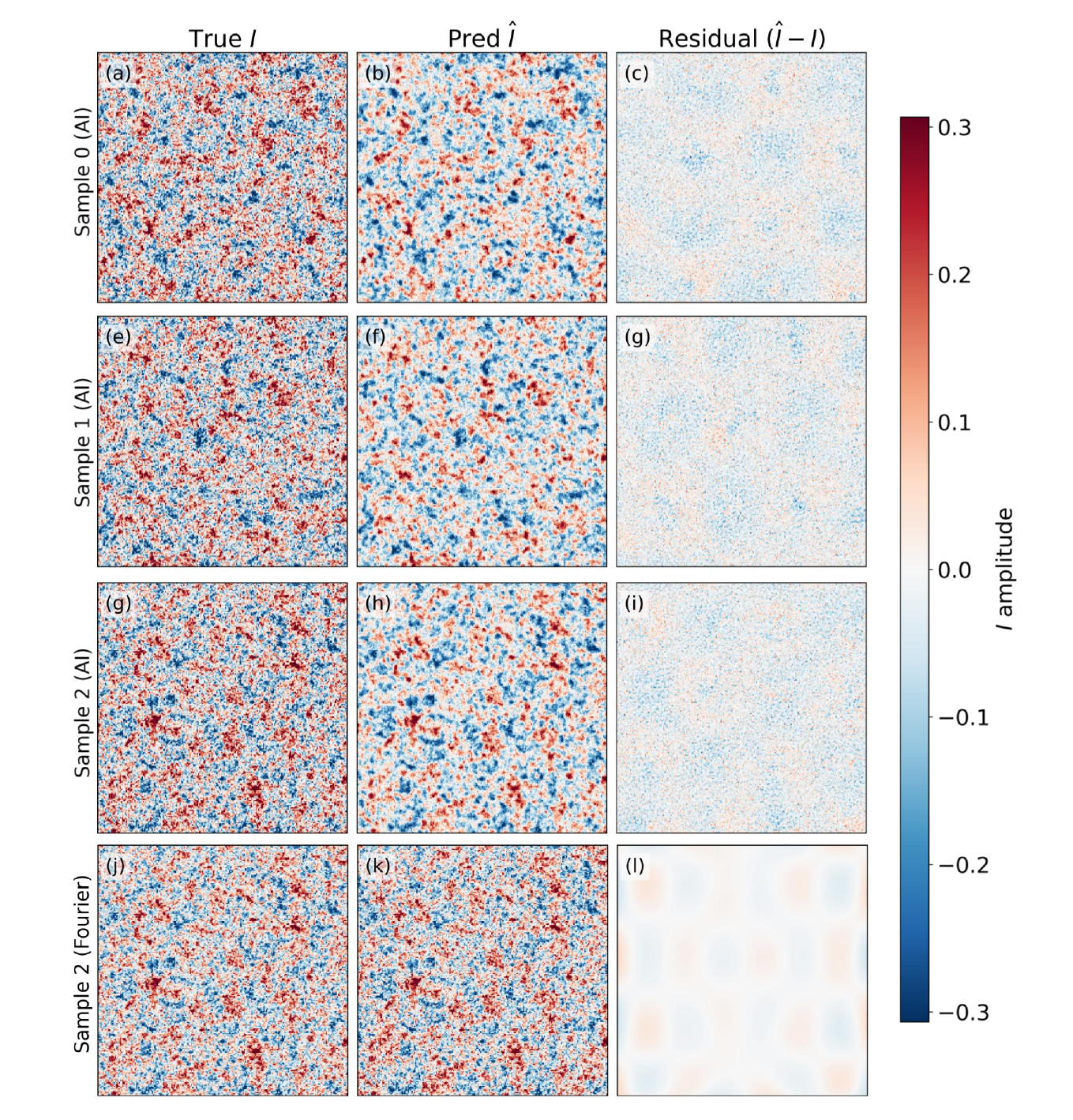}
    \caption{Examples of turbulence separation on the synthetic test dataset. The first three rows show the results obtained with the AI model for three independent test samples. Columns show the true turbulence field $I$, the predicted turbulence $\hat{I}$ from the AI model, and the residual $(\hat{I}-I)$. The residual maps indicate that the model accurately recovers the small-scale turbulent structures with only weak, noise-like errors. For comparison, the bottom row shows the result obtained using a standard Fourier filtering approach for the same sample. While Fourier filtering also captures a significant fraction of the turbulent fluctuations, coherent residual structures remain visible due to spectral overlap between the background flow and turbulence.}
    \label{fig:f4}
\end{figure*}

\begin{table}
\begin{tabular}{l c c c} 
 \hline
Model & $\mathcal{L}_{\rm MSE}$ & $\mathcal{L}_{\rm spec}$ & $|E_{\rm pred}-E_{\rm true}|/E_{\rm true}$\\ 
 \hline
 AI Model & 	$3.4 \times 10^{-4}$ & $2.6 \times 10^{-2}$ & $2.3 \times 10^{-4}$\\ 
 Fourier Filter & $2.1 \times 10^{-4}$ & $1.9 \times 10^{-2}$ & $2.0 \times 10^{-4}$\\ 
 \hline
\end{tabular}
\caption{
Quantitative comparison between the AI-based separation and a Fourier filtering approach on the synthetic test dataset.
The table reports the mean squared error ($\mathcal{L}_{\rm MSE}$), the spectral loss ($\mathcal{L}_{\rm spec}$), and the relative error in the turbulent energy budget,
$|E_{\rm pred}-E_{\rm true}|/E_{\rm true}$.
Both methods achieve comparable accuracy in reconstructing the turbulent component and preserving the total turbulent energy.
While Fourier filtering produces slightly smaller pixel-wise and spectral errors, the AI model yields a similarly accurate reconstruction without requiring an explicit spectral cutoff scale.
}
\label{Table2}
\end{table}

Figure~\ref{fig:f4} shows representative examples of turbulence separation on the synthetic test dataset.
The first three rows present the results obtained with the AI model for three independent test samples.
The predicted turbulence fields closely reproduce the spatial structure of the true turbulence,
while the residual maps remain small and largely noise-like, indicating that the model successfully isolates the turbulent component from the background field.
For reference, the bottom row shows the result obtained using a standard Fourier filtering approach for the same sample.
Both methods recover similar small-scale turbulent structures, indicating that the AI model performs comparably to traditional filtering techniques.
However, the Fourier filtering result exhibits coherent residual patterns associated with background leakage,which arise from spectral overlap between the background flow and turbulent fluctuations.

To further quantify the performance of the AI-based separation, we compare it with a standard Fourier filtering approach using the synthetic test dataset where the ground-truth turbulence field is known.
Table~\ref{Table2} summarizes the mean squared error, the spectral loss, and the relative error in the turbulent energy budget.
For the Fourier baseline, the turbulence field is obtained by applying a spectral high-pass filter that retains modes with $k \ge k_{\rm inj}$ while removing lower-wavenumber background modes.
The results show that the Fourier filtering method achieves slightly smaller pixel-wise and spectral errors.
However, both methods preserve the total turbulent energy with comparable accuracy.
This reflects the fact that Fourier filtering cannot perfectly separate background and turbulent modes when their spectral ranges partially overlap,
which can lead to residual background leakage into the filtered turbulence field.
In contrast, the AI model performs a data-driven separation that achieves a similarly accurate reconstruction of the turbulent component without requiring an explicit choice of a spectral cutoff scale.

\begin{figure}[t]
    \centering
    \includegraphics[width=1\linewidth]{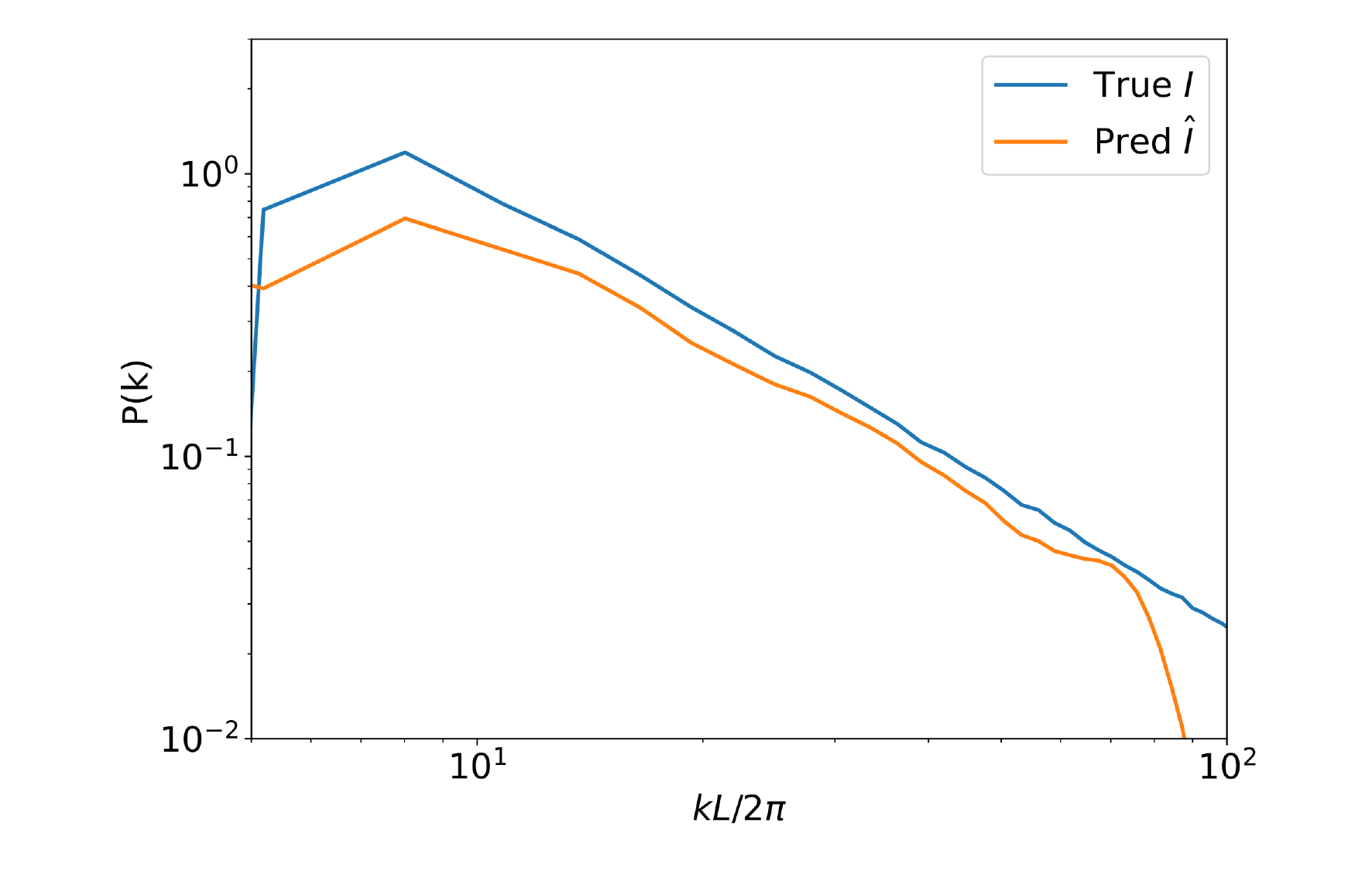}
    \caption{Comparison of the radially averaged power spectra for the true turbulence and the AI-predicted turbulence, averaged over 128 independent test samples. The predicted spectrum closely follows the true spectrum over a broad range of wavenumbers, including the inertial range, with deviations appearing only at the highest wavenumbers where finite-resolution and numerical dissipation effects become significant.}
    \label{fig:f5}
\end{figure}

\begin{table}
\begin{tabular}{l c c c} 
 \hline
Test Set & $\mathcal{L}_{\rm MSE}$ & $\mathcal{L}_{\rm spec}$ & $\mathcal{L}_{\rm leak}$ \\ 
 \hline
 baseline ($k^{-5/3}$) & 	$3.4 \times 10^{-4}$ & $2.6 \times 10^{-2}$ & $4.9 \times 10^{-3}$ \\ 
 $k^{-3/2}$& $4.1 \times 10^{-4}$ & $3.0 \times 10^{-2}$ & $5.2 \times 10^{-3}$ 	 \\ 
 $k^{-2}$& $3.5 \times 10^{-4}$ & $2.8 \times 10^{-2}$ & $5.1 \times 10^{-3}$ 	 \\
 \hline
\end{tabular}
\caption{Performance of the trained model when evaluated on synthetic test sets with different turbulent power spectra. 
Although the model is trained on a $k^{-5/3}$ spectrum, the separation accuracy remains broadly consistent for alternative slopes, including $k^{-3/2}$ and $k^{-2}$.}
\label{Table3}
\end{table}

To quantify the spectral accuracy of the separation, we compute power spectra for 128 independent test samples and average them. Fig.~\ref{fig:f5} compares the mean power spectrum of the true turbulence with that of the model prediction. 
The predicted spectrum closely follows the true spectrum over a broad range of wavenumbers, including the inertial range, with deviations appearing only at the highest wavenumbers. 
The discrepancy at the highest wavenumbers likely reflects a combination of factors. 
First, the reference turbulence fields themselves are subject to finite numerical resolution and dissipation, which naturally suppresses power near the grid scale. 
Second, neural network–based reconstructions are known to exhibit spectral bias, with a tendency to reproduce large-scale structures more accurately than small-scale fluctuations \cite[e.g.,][]{Fukami2019}. 
The combined effect of these factors likely contributes to the reduced agreement at the highest resolved wavenumbers.

To test the sensitivity of the method to the assumed turbulent spectrum, we evaluated the trained model on synthetic test sets generated with alternative power-law slopes, including $k^{-3/2}$ and $k^{-2}$. 
The results are summarized in Table~\ref{Table3}. The separation performance remains broadly consistent across these spectral variations, with only modest changes in the loss metrics. 
This indicates that the model is not strongly tied to the specific $k^{-5/3}$ spectrum used during training and retains good performance for moderate variations of the turbulent spectral slope.

\section{Application to Decaying Hydrodynamic Turbulence}
\label{sec:s3}

The AI model is trained exclusively on static synthetic data in which the background and turbulence components are linearly superposed and do not interact.
In contrast, physically realistic flows are governed by nonlinear dynamics, in which coherent structures and turbulence evolve self-consistently.
To assess the robustness and generalization capability of the model, we apply it to two-dimensional incompressible Navier--Stokes simulations of decaying hydrodynamic turbulence.

In this section, we primarily describe the numerical setup used to simulate decaying hydrodynamic turbulence in two dimensions.
The simulations are designed to provide a controlled physical testbed in which an initially imposed large-scale coherent flow interacts nonlinearly with freely decaying turbulence, a setting that has been widely used to study the emergence of coherent structures and late-time organization in two-dimensional incompressible flows \citep[e.g.,][]{Kraichnan1980,Matthaeus1991,Dmitruk1996}.
No external forcing is applied after the initial time, allowing us to isolate the intrinsic evolution of turbulence and its interaction with the background flow.
This setup is particularly suited for assessing the robustness of AI-based turbulence--background separation models under time-evolving flow conditions.
We then apply a neural network trained on static synthetic data to snapshots of the evolving simulations and analyze its ability to recover turbulent fluctuations through a combination of spectral and energy-based diagnostics.

\subsection{Governing Equations}

We simulate two-dimensional incompressible turbulence by solving the Navier--Stokes equations in vorticity--streamfunction form.
Under the incompressibility condition,
\begin{equation}
\nabla \cdot \mathbf{u} = 0,
\end{equation}
the evolution of the vorticity $\omega$ is governed by
\begin{equation}
\frac{\partial \omega}{\partial t}
+ \mathbf{u} \cdot \nabla \omega
= \nu \nabla^2 \omega + f,
\end{equation}
where $\nu$ is the kinematic viscosity and $f$ represents an external forcing term.  
In the decaying turbulence experiments discussed in this work, we set $f=0$.

In two dimensions, the velocity field $\mathbf{u}=(u,v)$ is expressed in terms of the streamfunction $\psi$ as
\begin{equation}
u = \frac{\partial \psi}{\partial y}, \qquad
v = -\frac{\partial \psi}{\partial x},
\end{equation}
and the vorticity is given by
\begin{equation}
\omega = \nabla^2 \psi.
\end{equation}

At each time step, the Poisson equation for $\psi$ is solved in Fourier space to recover the velocity field from the vorticity field.

\subsection{Computational Domain and Non-dimensionalization}

The simulations are performed in a doubly periodic square domain,
\begin{equation}
(x,y) \in [0,L_x) \times [0,L_y),
\end{equation}
with $L_x=L_y=1$.  
All quantities are expressed in non-dimensional units.  
Consequently, the simulation time $t$ should be interpreted as a dimensionless time measured in units of the characteristic flow crossing time, $\tau \sim L/U$, where $U$ is a representative velocity scale of the system.

\subsection{Pseudo-spectral Method and De-aliasing}

Spatial derivatives are computed using a pseudo-spectral method based on Fourier transforms.  
In Fourier space, derivatives are evaluated exactly via
\begin{equation}
\partial_x \rightarrow i k_x, \qquad
\partial_y \rightarrow i k_y, \qquad
\nabla^2 \rightarrow -k^2,
\end{equation}
where $k^2 = k_x^2 + k_y^2$.

The streamfunction in Fourier space is obtained from the vorticity spectrum as
\begin{equation}
\hat{\psi}(\mathbf{k}) = -\frac{\hat{\omega}(\mathbf{k})}{k^2}, \qquad (k \neq 0),
\end{equation}
with the zero mode set to zero.  
The velocity components are then reconstructed via
\begin{equation}
\hat{u} = i k_y \hat{\psi}, \qquad
\hat{v} = -i k_x \hat{\psi}.
\end{equation}

The nonlinear advection term $\mathbf{u}\cdot\nabla\omega$ is evaluated in real space following the standard pseudo-spectral procedure.  
To suppress aliasing errors arising from nonlinear mode coupling, the $2/3$ de-aliasing rule is applied by truncating Fourier modes whose wavenumbers exceed two-thirds of the maximum resolvable wavenumber.

\subsection{Time Integration}

Time integration is performed using a second-order Runge--Kutta (RK2) scheme.  
Given the right-hand side operator $R(\omega)$, the update from time level $n$ to $n+1$ reads
\begin{equation}
\omega^{(1)} = \omega^n + \Delta t\, R(\omega^n),
\end{equation}
\begin{equation}
\omega^{n+1} = \omega^n + \frac{\Delta t}{2}
\left[ R(\omega^n) + R(\omega^{(1)}) \right].
\end{equation}
De-aliasing is applied after each sub-step.

\subsection{Initial Conditions}

\subsubsection{Coherent Background Flow}

The large-scale background flow is constructed using a streamfunction of the form
\begin{equation}
\psi_{\rm bg}(x,y)
= A_{\psi}
\sin(2\pi k_x x + \phi_1)
\cos(2\pi k_y y + \phi_2),
\end{equation}
where $k_x$ and $k_y$ denote the number of cycles across the domain, and $\phi_1$, $\phi_2$ are random phases.

A key implementation detail is that the parameter controlling the background amplitude is defined in terms of the target velocity root-mean-square (RMS),
\begin{equation}
u_{\rm rms,bg}
=
\left[
\frac{1}{2}
\left(
\langle u^2 \rangle + \langle v^2 \rangle
\right)
\right]^{1/2},
\end{equation}
rather than the streamfunction amplitude $A_{\psi}$.  
After constructing the background streamfunction, the entire field is rescaled so that the resulting velocity field attains the prescribed RMS value.  
This normalization ensures numerical stability and provides a physically intuitive control parameter.

\subsubsection{Turbulent Perturbations}

Divergence-free turbulent perturbations are added to the background flow at the initial time.  
The turbulent component is generated in Fourier space using a random-phase streamfunction spectrum designed to approximate a target energy spectrum,
\begin{equation}
E(k) \propto k^{s},
\end{equation}
with $s=-5/3$ in this study.  
The perturbations are confined to a finite injection band centered around a characteristic wavenumber $k_{\rm inj}$.

The initial turbulent velocity amplitude is specified relative to the background RMS via
\begin{equation}
u_{\rm rms,turb}(t=0)
=
\texttt{turb\_ratio}
\times
u_{\rm rms,bg}.
\end{equation}

The total initial vorticity field is therefore
\begin{equation}
\omega(t=0)
=
\omega_{\rm bg}(t=0)
+
\omega_{\rm turb}(t=0).
\end{equation}

\begin{figure*}[t]
    \centering
    \includegraphics[width=1\linewidth]{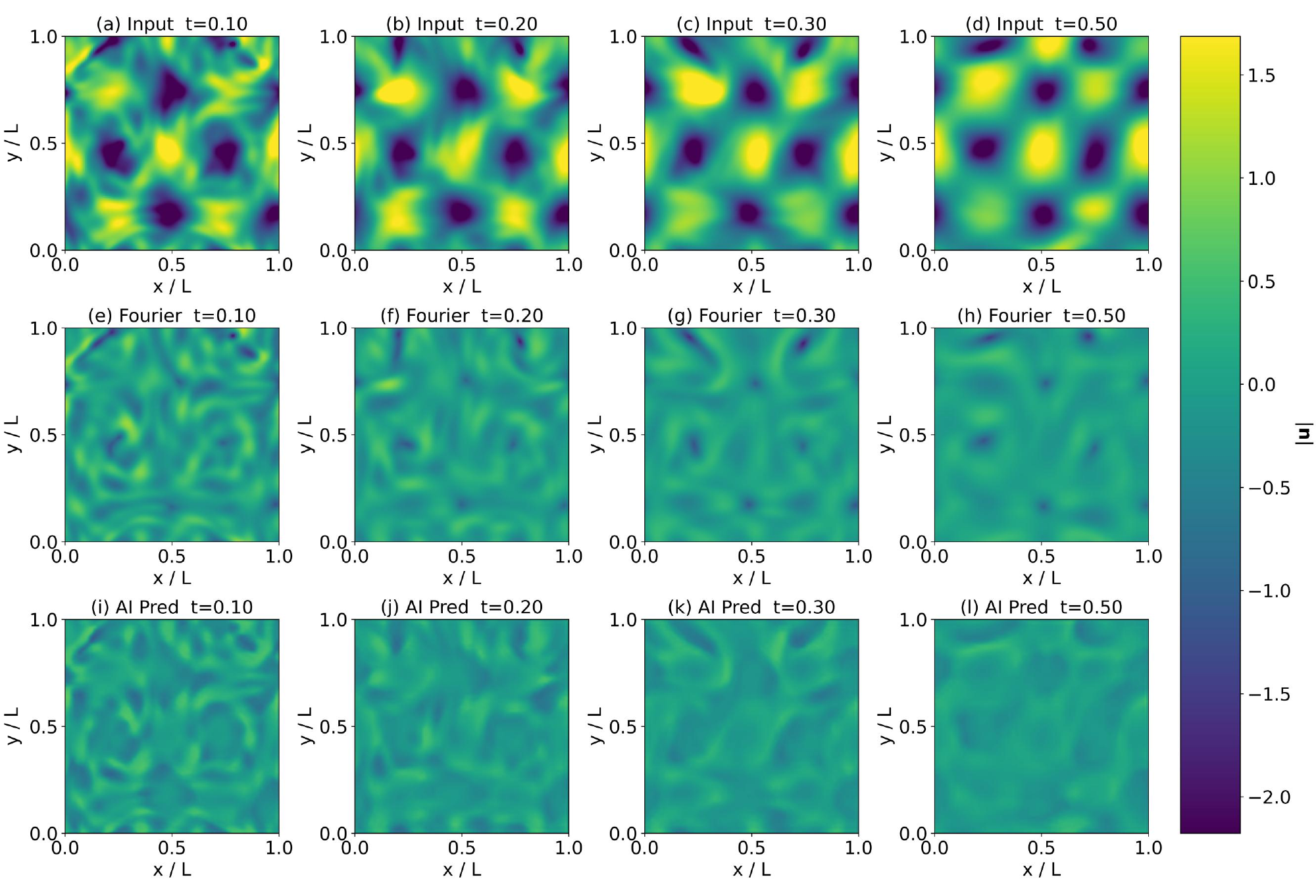}
    \caption{Application of the trained AI model to snapshots from a two-dimensional incompressible Navier-Stokes simulation of decaying turbulence. The top row shows the original simulation fields containing both large-scale background flow and turbulent fluctuations. The middle row shows turbulence extracted using a Fourier filtering approach, and the bottom row shows the turbulence predicted by the AI model. Because the background and turbulence interact nonlinearly during the simulation, a unique reference turbulence field is not available.}
    \label{fig:f6}
\end{figure*}

\begin{figure}[t]
    \centering
    \includegraphics[width=1\linewidth]{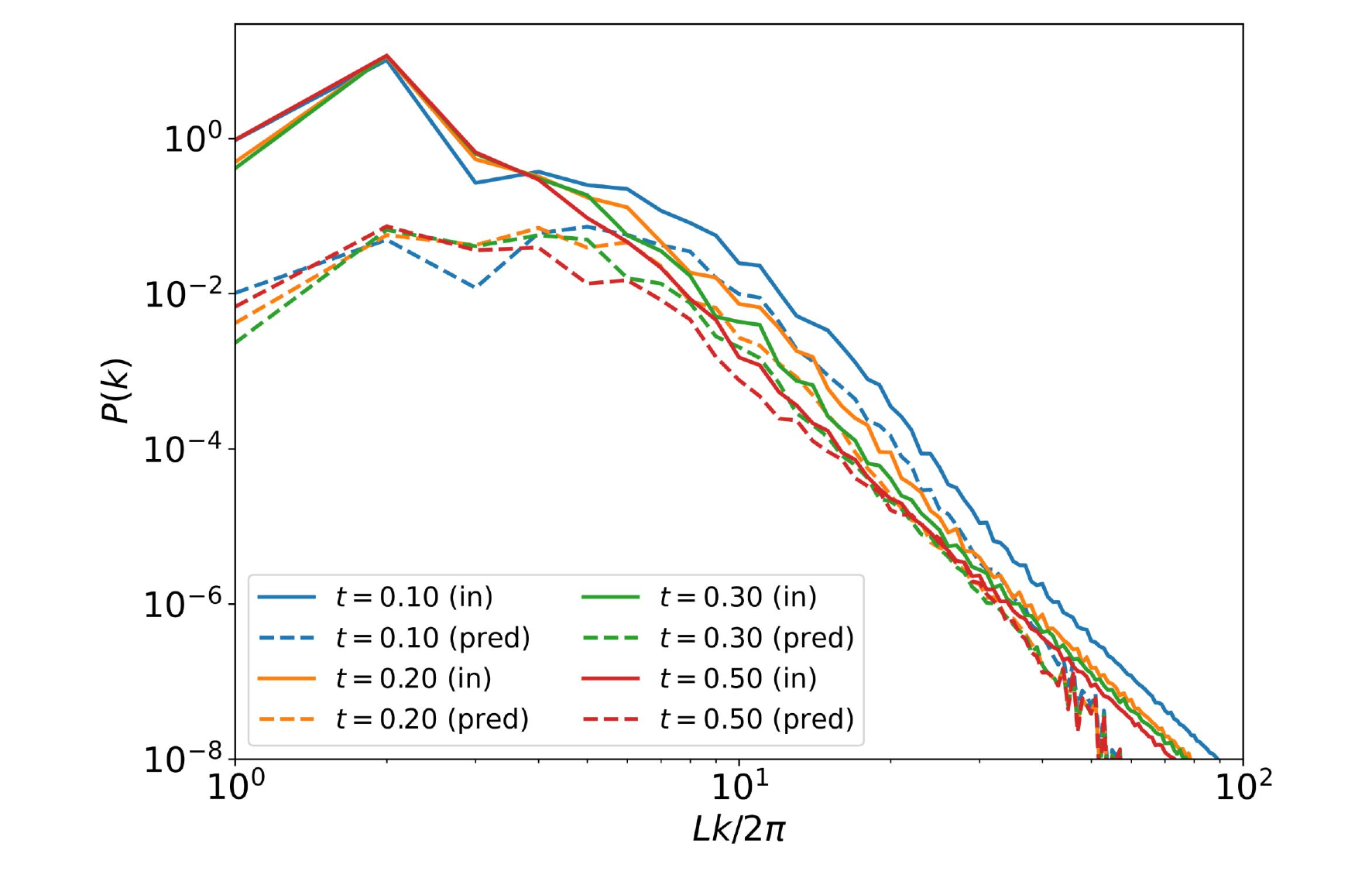}
    \caption{Power spectra of the original simulation fields and the AI-extracted turbulence at multiple simulation times. The removal of large-scale power at low wavenumbers is clearly visible in the AI-processed fields, while the inertial-range scaling at higher wavenumbers is largely preserved. This demonstrates that the AI model performs an effective scale-selective separation consistent with the physical notion of turbulence, even in a time-evolving nonlinear system. No retraining or fine-tuning was performed on the model when applied to the simulation data shown here.}
    \label{fig:f7}
\end{figure}

\begin{figure}[t]
    \centering
    \includegraphics[width=1\linewidth]{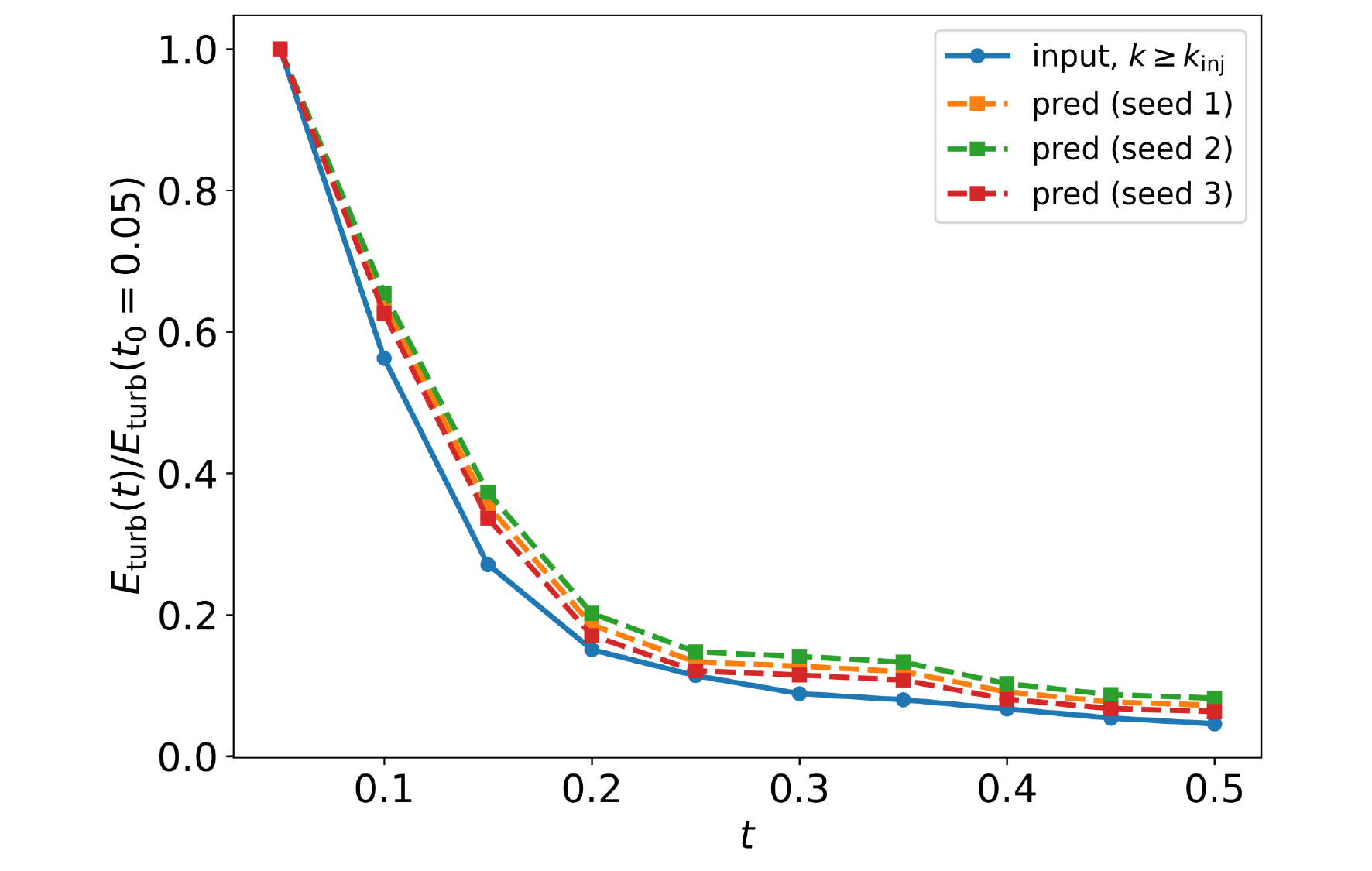}
    \caption{Time evolution of the turbulent-band energy inferred from the input field (solid line) and the turbulence fields recovered by three independently trained AI models with different random initializations (dashed lines). The turbulent energy is defined as $E_{\rm turb}(t)=\sum_{k\ge k_{\rm inj}} P(k,t)$ with $Lk_{\rm inj}/2\pi=8$, and is normalized by its value at the earliest available time ($t_0=0.05$). Restricting the integration to $k\ge k_{\rm inj}$ isolates the turbulent contribution. The nearly overlapping AI curves indicate that the recovered energy evolution is robust to variations in model initialization.}
    \label{fig:f8}
\end{figure}

\subsection{Simulation Parameters and Stability Monitoring}

The simulations presented in this work adopt the following parameters:
\begin{itemize}
\item Grid resolution: $N_x = N_y = 224$
\item Domain size: $L_x = L_y = 1$
\item Kinematic viscosity: $\nu = 2\times10^{-3}$
\item Time step: $\Delta t = 10^{-4}$
\item Total number of steps: $10000$
\item Output cadence: every 500 steps
\item Background RMS velocity: $u_{\rm rms,bg} = 1.0$
\item Turbulence-to-background ratio: $\texttt{turb\_ratio} = 0.2$
\item Injection scale: $Lk_{\rm inj}/2\pi = 8$
\end{itemize}

No external forcing is applied in the decaying turbulence runs.
Each snapshot stores the velocity field $(u,v)$, the vorticity $\omega$, and the isotropically averaged energy spectrum, enabling detailed analyses of spectral evolution, background--turbulence interaction, and AI-based separation performance.

Numerical stability is monitored through a CFL-like condition,
\begin{equation}
\mathrm{CFL} \approx
\frac{u_{\max} \Delta t}{\Delta x},
\end{equation}
where $u_{\max}$ is the maximum velocity magnitude at a given time and $\Delta x = L_x / N_x$.  
In all simulations, the CFL number remains well below unity, ensuring stable time integration.

\subsection{Results}

The present simulation is not intended as a fully realistic model of astrophysical turbulence, which is typically continuously forced.  
Instead, it provides a controlled physical testbed in which a coherent large-scale background flow interacts nonlinearly with decaying turbulence.  
This setup is particularly well suited for assessing the robustness and generalization of AI-based methods trained on static snapshots when applied to time-evolving, physically interacting flow fields.
Although the simulation is two-dimensional, the controlled nature of the setup allows us to isolate the effects of nonlinear scale interaction on the separation task.

In the absence of external forcing, the turbulent energy decays monotonically with time, and the distinction between turbulent fluctuations and coherent large-scale structures becomes progressively less well defined.
At sufficiently late times, the turbulent component is energetically subdominant and strongly entangled with the evolving background flow, rendering any turbulence--background separation intrinsically ambiguous.
For this reason, we focus our analysis on the interval $t\le 0.5$, during which the turbulent band remains dynamically significant and the separation problem is physically meaningful.
Beyond this stage, further degradation of the separation reflects the intrinsic ambiguity of the problem rather than a change in the behavior or robustness of the AI model.

Fig.~\ref{fig:f6} shows snapshots of the hydrodynamic simulation at several times,
together with turbulence fields extracted using both a Fourier filtering approach
and the AI-based model.
The top row displays the original simulation fields containing both the coherent
large-scale background flow and small-scale turbulent fluctuations.
Because the background and turbulent components interact nonlinearly during the
simulation, a unique reference turbulence field is not available at later times.
As a result, the separation results can only be evaluated qualitatively.
The middle row shows the turbulence fields obtained using a standard Fourier
filtering method, while the bottom row shows the turbulence extracted by the
AI model.
At early times, both methods recover small-scale turbulent structures that are
visually consistent with those present in the input flow.
As the simulation evolves and nonlinear interactions increasingly distort the
background flow, the separation becomes more ambiguous.
Nevertheless, the AI model continues to extract a physically plausible turbulent
component that remains comparable to the result obtained with the Fourier filtering
approach.

It is important to note that the training data used in this study consist of linear superpositions of background and turbulent components. 
In contrast, the hydrodynamic simulations shown here involve nonlinear interactions between these components as the flow evolves. 
This mismatch is intentional and serves as a test of whether a model trained only to identify coherent background patterns can still be meaningfully applied to flow fields whose detailed turbulence evolution is governed by more complex nonlinear physics. 
The results indicate that the model remains effective while coherent and turbulent structures remain reasonably distinguishable, whereas the separation becomes increasingly ambiguous as nonlinear coupling distorts that distinction at later times. 
In the training procedure, the leakage loss term uses the known background component available in the synthetic dataset to discourage contamination of the extracted turbulence by large-scale modes. 
This constraint is applied only during training and is not required during inference, so the trained model can be applied to arbitrary input fields without knowledge of the true background.

The scale-dependent performance of the turbulence--background separation is quantified in Fig. \ref{fig:f7}, which compares the power spectra of the original simulation fields and the AI-extracted turbulence at multiple times.
In all cases, the removal of large-scale power at low wavenumbers is clearly visible, while the inertial-range scaling at higher wavenumbers is largely preserved.
This demonstrates that the AI model performs an effective large-scale filtering that is consistent with the physical notion of turbulence, even in a time-evolving nonlinear system.

As a complementary and physically transparent diagnostic, we examine the time evolution of the turbulent-band energy inferred from the input and AI-recovered fields.
The turbulent energy is defined as
\begin{equation}
E_{\rm turb}(t)=\sum_{k\ge k_{\rm inj}} P(k,t) 2\pi k\Delta k,
\end{equation}
with $Lk_{\rm inj}/2\pi=8$, and is normalized by its value at the earliest available time.
Although the input field contains both coherent and turbulent components, restricting the spectral integration to $k\ge k_{\rm inj}$ ensures that the measured energy predominantly traces the turbulent contribution.
As shown in Fig.~\ref{fig:f8}, the turbulent energy extracted from three independently trained AI models with different random initializations exhibits a decay trend that closely follows that of the input field over the entire interval $t\le0.5$.
All three AI curves remain very similar to each other, indicating that the recovered turbulent-energy evolution is robust to stochastic variations in model initialization.
Both the input and AI-recovered curves show a rapid early-time decay followed by a more gradual decline at later times, indicating that the model captures the overall temporal evolution of the turbulent energy despite increasing spectral overlap between coherent and turbulent structures.
The recovered energy remains slightly higher than the input at late times, which is consistent with residual contamination from large-scale components and does not indicate a qualitative mismatch in the decay behavior.
These results demonstrate that, while the separation becomes increasingly ambiguous at later stages, the AI model robustly tracks the global energy decay of the turbulent band in a physically meaningful manner.

\section{Summary and Discussion}
\label{sec:s4}

In this work, we have investigated the robustness of an AI-based turbulence--background separation model when applied to physically time-evolving hydrodynamic flows.
The model was trained exclusively on static synthetic data constructed as linear superpositions of a coherent background component and a turbulent fluctuation field.
Despite this highly idealized training setup, we demonstrated that the model can successfully extract turbulent structures from snapshots of two-dimensional incompressible Navier--Stokes simulations in which the background flow and turbulence evolve self-consistently and interact nonlinearly.
This result indicates that the AI model is capable of generalizing beyond the specific statistical properties of the training data, provided that a meaningful physical distinction between coherent and turbulent scales is retained.

A key aspect of this study is the use of decaying hydrodynamic turbulence as a testbed.
Unlike forced turbulence, decaying turbulence represents a conservative scenario in which no additional energy is injected and the separation between large-scale and small-scale structures becomes progressively more challenging over time.
As the turbulent energy decays and nonlinear interactions distort the initially coherent background flow, the amplitude of turbulent fluctuations decreases and the distinction between background and turbulence becomes increasingly subtle.
Nevertheless, the AI model continues to recover visually and spectrally plausible small-scale turbulent structures, even when the turbulent energy has significantly decayed, providing a stringent test of its robustness under adverse conditions.

Our results also provide insight into what the AI model has learned.
The successful recovery of turbulent fluctuations does not appear to arise from a simple high-pass filtering or smoothing operation.
Instead, the preservation of inertial-range spectral scaling, together with the suppression of low-wavenumber power associated with the background flow, suggests that the model has learned a scale-selective representation of turbulence that is broadly consistent with its physical definition.
This interpretation is further supported by the fact that the model maintains reasonable performance even when the background flow is substantially deformed relative to the training examples, indicating that the separation relies on more than purely morphological pattern matching.
Consistent with this interpretation, quantitative comparisons with a traditional Fourier filtering approach show that the AI model achieves comparable reconstruction accuracy in terms of pixel-wise error, spectral consistency, and preservation of the turbulent energy budget.
This result suggests that the learned separation performs at a level similar to conventional scale-based filtering while not requiring an explicit choice of a spectral cutoff scale.

The temporal evolution of the simulation data allows us to identify a natural failure boundary for the AI-based separation.
At sufficiently late times (i.e., $t > 0.5$), when nonlinear interactions lead to strong spectral overlap between background and turbulent components, we interpret the degradation of the AI output as reflecting an intrinsic physical ambiguity rather than a deficiency of the model architecture or training procedure.
Explicitly identifying this temporal boundary is an important outcome of the present study, as it clarifies the conditions under which AI-based turbulence--background separation can be meaningfully applied.

Although the present work is limited to two-dimensional incompressible flows and decaying turbulence, the approach provides a clear framework for future extensions.
We note that two-dimensional turbulence differs in important ways from fully three-dimensional turbulence, in particular through the tendency to sustain stronger large-scale coherent structures and different cascade dynamics. As a result, the separation problem considered here may represent a somewhat simplified scenario compared to realistic astrophysical turbulence.
Nevertheless, the use of decaying two-dimensional turbulence provides a controlled and computationally tractable benchmark in which the transition from idealized static superpositions to self-consistent nonlinear flow evolution can be examined in a transparent manner. Within this context, the present setup allows us to evaluate whether AI models trained on static synthetic data can generalize to dynamically evolving turbulent systems.
Natural next steps include the application to forced turbulence, fully three-dimensional flows, and compressible systems more directly relevant to astrophysical environments. In addition, incorporating temporal information explicitly into the AI architecture may further improve robustness in strongly time-dependent regimes.
Overall, this study demonstrates that decaying hydrodynamic turbulence offers a physically meaningful and conservative benchmark for evaluating AI-based turbulence--background separation, and highlights both the potential and the limitations of applying deep-learning models trained on static data to dynamically evolving turbulent systems.
We therefore emphasize that AI-based separation should not be interpreted as defining a unique physical decomposition, but rather as providing a data-driven diagnostic whose validity depends on the underlying flow regime.

\section*{Acknowledgements}
The authors thank the referees for their comments and suggestions, which have improved the quality of this manuscript.

\appendix
\setcounter{table}{0}
\renewcommand{\thetable}{A.\arabic{table}}

\section{Sensitivity to Model Architecture}
\label{sec:sa1}

\begin{table*}[h]
\centering
\caption{Comparison of model variants with different architectural configurations. The table reports the number of trainable parameters and evaluation metrics on the synthetic test dataset.}
\label{tab:A1}
\begin{tabular}{lcccc}
\hline
Model & Parameters (million) & $\mathcal{L}_{\rm MSE}$ & $\mathcal{L}_{\rm spec}$ & $\mathcal{L}_{\rm leak}$ \\
\hline
Baseline & 35.9 & $3.4 \times 10^{-4}$ & $2.6 \times 10^{-2}$ & $4.9 \times 10^{-3}$ \\
Deeper decoder & 34.9 & $3.6 \times 10^{-4}$ & $2.7 \times 10^{-2}$ & $5.1 \times 10^{-3}$ \\
Reduced decoder & 28.7 & $3.7 \times 10^{-4}$ & $2.9 \times 10^{-2}$ & $5.3 \times 10^{-3}$ \\
\hline
\end{tabular}
\end{table*}

To assess the sensitivity of the results to the architectural design of the
neural network, we performed two additional controlled experiments. In these
tests, we kept the Swin-Tiny encoder fixed and modified only the decoder
architecture, allowing us to isolate the impact of decoder design while
preserving the learned feature representation of the encoder.

The baseline decoder uses two 3 × 3 convolutional layers per block
with channel sizes $\{64, 128, 256, 512\}$.
In the first variant, the number of convolutional layers per block
is increased from 2 to 3, while the channel sizes are reduced to
$\{56, 112, 224, 448\}$ to maintain a comparable number of trainable
parameters.
In the second variant, the decoder structure is kept identical to
the baseline (two convolutional layers per block), but the channel
sizes are reduced to $\{32, 64, 128, 256\}$, resulting in a smaller
number of trainable parameters. In both cases, the Swin-Tiny encoder was kept fixed,
and only the decoder architecture was modified.
All models were trained using the same dataset, loss function, optimizer, and
number of training epochs as the baseline model. 

The results are summarized in Table~\ref{tab:A1}.
We find that increasing the compositional capacity of the decoder at a
comparable parameter count does not lead to a significant improvement in
performance. The reduced-decoder model retains broadly comparable reconstruction
accuracy, although with a modest degradation in both spectral consistency and
leakage suppression.
We note that reducing only the decoder width does not halve the total number of
trainable parameters, as the majority of parameters reside in the Swin-Tiny
encoder. Nevertheless, this variant provides a controlled test of reduced
decoder capacity while preserving the overall architecture and keeping the
encoder fixed.
Overall, these results indicate that the model performance is not strongly
sensitive to moderate variations in the decoder design, and that the baseline
configuration provides a stable and robust choice for the present task.

\bibliographystyle{elsarticle-harv} 
\bibliography{reference}



%% For example:

%% \bibitem[Aladro et al.(2015)]{Aladro15} Aladro, R., Martín, S., Riquelme, D., et al. 2015, \aas, 579, A101

%%\end{thebibliography}

\end{document}